\documentclass{svmult}

\usepackage{makeidx}         
\usepackage{graphicx}        
\usepackage{multicol}        
\usepackage[bottom]{footmisc}

\makeindex             

\newcommand{\pd}{\partial}             
\newcommand{\la}{\left\langle}         
\newcommand{\ra}{\right\rangle}        
\newcommand{\mr}[1]{{\mathrm {#1}}}    
\newcommand{\kB}{k_\mr{B}}             
\newcommand{\eq}[1]{Eq.~(\ref{#1})}    
\newcommand{\eqs}[1]{Eqs.~(\ref{#1})}  
\newcommand{\rf}[1]{Ref.~\cite{#1}}    
\newcommand{\rfs}[1]{Refs.~\cite{#1}}  
\newcommand{\fg}[1]{Fig.~\ref{#1}}     
\newcommand{\fgs}[1]{Figs.~\ref{#1}}   

\begin{document}

\title*{Kinetic theory models for the distribution of wealth:
power law from overlap of exponentials}
\titlerunning{Kinetic theory models for the distribution of wealth}
\author{
  Marco Patriarca\inst{1,3}
  \and
  Anirban Chakraborti\inst{2}
  \and
  Kimmo Kaski\inst{1}
  \and
  Guido~Germano\inst{3}
}
\authorrunning{Patriarca, Chakraborti, Kaski, Germano}

\institute{
 Laboratory of Computational Engineering,
 Helsinki University of Technology,
 POBoX 9203, 02015 HUT, Finland
 \and
 Brookhaven National Laboratory,
 Department of Physics, Upton,
 New York 11973, USA
 \and
 Fachbereich Chemie, Philipps-Universit\"at Marburg,
 35032 Marburg, Germany
 \vspace{0.25cm}
 \\ \texttt{patriarc@staff.uni-marburg.de}
 \\ \texttt{anirban@bnl.gov}
 \\ \texttt{kimmo.kaski@hut.fi}
 \\ \texttt{germano@staff.uni-marburg.de}
}

%
%
\maketitle

\graphicspath{{./fig/}}  

\begin{abstract}
Various multi-agent models of wealth distributions defined by microscopic
laws regulating the trades, with or without a saving criterion, are reviewed.
We discuss and clarify the equilibrium properties of the model with constant
global saving propensity, resulting in Gamma distributions, and their
equivalence to the Maxwell-Boltzmann kinetic energy distribution for
a system of molecules in an effective number of dimensions $D_\lambda$,
related to the saving propensity $\lambda$ 
[M. Patriarca, A. Chakraborti, and K. Kaski, Phys. Rev. E 70 (2004) 016104].
We use these results to analyze
the model in which the individual saving propensities of the agents are
quenched random variables, and the tail of the equilibrium wealth distribution
exhibits a Pareto law $f(x) \propto x^{-\alpha -1}$ with an exponent $\alpha=1$
[A. Chatterjee, B. K. Chakrabarti, and S. S. Manna, Physica Scripta T106
(2003) 367]. Here, we show that the observed Pareto
power law can be explained as arising from the overlap of the
Maxwell-Boltzmann distributions associated to the various agents,
which reach an equilibrium  state characterized by their individual Gamma
distributions.
We also consider the influence of different types of saving propensity
distributions on the equilibrium state.
\end{abstract}

\section{Introduction}
\label{sec:intro}

\begin{quote}
{\it `A rich man is nothing but a poor man with money'} --- W. C. Fields.
\end{quote}

If money makes the difference in this world, then it is perhaps wise to
dwell on what money, wealth and income are,
to study models for predicting the respective distributions,
how they are divided among the population of a given country
and among different countries.
The most common definition of {\em money} suggests that money is the
``Commodity accepted by general consent as medium of economics exchange''
\cite{eb}.
In fact money circulates from one economic agent (which can be
an individual, firm, country, etc.) to another, thus facilitating trade.
It is ``something which all other goods or services are traded for''
(for details see \rf{Shostak2000a}) .
Throughout history various commodities have been used as money, termed usually
as ``commodity money'' which include rare seashells or beads, and cattle (such
as cows in India). Since the 17th century the most common forms have been
metal coins, paper notes, and book-keeping entries.
However, this is not the only important point about money.
It is worth recalling the four functions of money
according to standard economic theory:

\begin{enumerate}
\item to serve as a medium of exchange universally accepted in trade for goods
and services
\item to act as a measure of value, making possible the determination of the
prices and the calculation of costs, or profit and loss
\item to serve as a standard of deferred payments, i.e., a tool for the payment
of debt or the unit in which loans are made and future transactions are fixed
\item to serve as a means of storing wealth not immediately required for use.
\end{enumerate}
A main feature that emerges from these properties
and that is relevant from the point of view
of the present investigation is that money
is the medium in which prices or values
of all commodities as well as costs, profits, and transactions
can be determined or expressed.
As for the {\em wealth},
it usually refers to things that have economic utility
(monetary value or value of exchange), or material goods or property.
It also represents the abundance of objects of value (or riches)
and the state of having accumulated these objects.
For our purpose, it is important to bear in mind that
wealth can be measured in terms of money.
Finally, {\em income} is defined as
``The amount of money or its equivalent received during a period of time in
exchange for labor or services, from the sale of goods or property, or as
profit from financial investments'' \cite{answers}.
Therefore, it is also a quantity which can be measured in terms
of money (per unit time).
Thus, money has a two-fold fundamental role, as
(i) an exchange medium in economic transactions, and
(ii) a unit of measure which allows one to quantify (movements
of) any type of goods which would otherwise be ambiguous to estimate.
The similarity with e.g., thermal energy
(and thermal energy units) in physics is to be noticed.
In fact, the description of the mutual transformations
of apparently different forms of energy,
such as heat, potential and kinetic energy, is made possible by the
recognition of their equivalence and the corresponding
use of a same unit. And it so happens that this same unit
is also the traditional unit used for one of the forms of energy. For example,
one could measure energy in all its forms, as actually done in some fields
of physics, in degree Kelvin.
Without the possibility of expressing different goods
in terms of the same unit of measure,
there simply would not be any quantitative approach to economy models,
just as there would be no quantitative description
of the transformation of the heat in motion and vice versa,
without a common energy unit.

\section{Multi-agent models for the distribution of wealth}
\label{sec:models}

In recent years several works have considered multi-agent models
of a closed economy
\cite{Bennati1988a,Bennati1988b,Bennati1993a,Dragulescu2000a,Chakraborti2000a,Chakraborti2002a,Chatterjee2003a,Chatterjee2004a,Chatterjee2005a}.
Despite their simplicity, these models predict a realistic shape of
the wealth distribution, both in the low income part, usually
described by a Boltzmann (exponential) distribution, as well in the tail,
where a power law was observed a century ago by the Italian
social economist Pareto \cite{Pareto1897a}: the wealth of
individuals in a stable economy follows the
distribution, $F(x)\propto x^{-\alpha }$, where $F(x)$ is
the upper cumulative distribution function, that is the number
of people having wealth greater than or equal to $x$, and $\alpha $
is an exponent (known as the Pareto exponent) estimated
to be between $1$ and $2$.
In such models, $N$ agents exchange a quantity $x$, that
has sometimes been defined as wealth and other times as money.
As explained in the introduction, here money must be interpreted
all the goods that constitute the agents' wealth expressed in the
same currency. To avoid confusion, in the following we will use
only the term wealth.
The states of agents are characterized by the wealths
$\{x_n\},~n=1,2,\dots,N$.
The evolution of the system is then carried out according to a prescription,
which defines a ``trading rule'' between agents.
The evolution can be interpreted both as an actual time evolution
or a Monte Carlo optimization procedure, aimed at finding
the equilibrium distribution.
At every time step two agents $i$ and $j$ are extracted at random
and an amount of wealth $\Delta x$ is exchanged between them,
\begin{eqnarray}
  x_i' &=& x_i - \Delta x \, ,
  \nonumber \\
  x_j' &=& x_j + \Delta x \, .
  \label{basic0}
\end{eqnarray}
It can be noticed that in this way the quantity $x$ is conserved
during the single transactions, $x_i'+x_j' = x_i + x_j$.
Here $x_i'$ and $x_j'$ are the agent wealths
after the transaction has taken place.
Several rules have been studied for the model.

\subsection{Basic model without saving: Boltzmann distribution}
\label{sec:basic}

In the first version of the model,
so far unnoticed in later literature,
the money difference $\Delta x$
is assumed to have a constant value
\cite{Bennati1988a,Bennati1988b,Bennati1993a},
\begin{equation} \label{B}
  \Delta x = \Delta x_0 \ .
\end{equation}
This rule, together with the constraint
that transactions can take place only
if $x_i'>0$ and $x_j'>0$, provides a Boltzmann distribution;
see the curve for $\lambda=0$ in \fg{fig:gamma}.
An equilibrium distribution with exponential tail is also obtained
if $\Delta x$ is a random fraction $\epsilon$
of the wealth of one of the two agents, 
$\Delta x = \epsilon x_i$ or $\Delta x = \epsilon x_j$.
A trading rule based on the random
redistribution of the sum of the wealths of the two agents
has been introduced by Dragulescu and Yakovenko \cite{Dragulescu2000a},
\begin{eqnarray}
  x_i' &=& \epsilon (x_i + x_j) \, ,
  \nonumber \\
  x_j' &=& \bar{\epsilon} (x_i + x_j) \, ,
  \label{basic1}
\end{eqnarray}
where $\epsilon$ is a random number uniformly distributed between 0 and 1
and $\bar{\epsilon}$ is the complementary fraction,
i.e.\ $\epsilon + \bar{\epsilon}  = 1$.
Equations (\ref{basic1}) are easily shown to correspond 
to the trading rule (\ref{basic0}), with
\begin{equation} \label{DY}
  \Delta x =  \bar{\epsilon} x_i - \epsilon x_j   \ .
\end{equation}
In the following we will concentrate on the latter version of the model
and its generalizations, though both the versions of the basic model 
defined by \eqs{B} or (\ref{DY}) lead to the Boltzmann distribution,
\begin{equation}
  f(x) = \frac{1}{ \la x \ra }\exp\left(-\frac{x}{\la x \ra}\right) \, ,
  \label{BD}
\end{equation}
where the effective temperature of the system is just
the average wealth $\la x \ra$
\cite{Bennati1988a,Bennati1988b,Bennati1993a,Dragulescu2000a}.
The result (\ref{BD}) is found to be robust,
in that it is largely independent of various factors.
Namely, it is obtained for the various forms of $\Delta x$
mentioned above, for pairwise as well as multi-agent interactions,
for arbitrary initial conditions \cite{Chakraborti2000a},
and finally, for random or consecutive extraction of the interacting
agents.
The Boltzmann distribution thus obtained has been sometimes referred to as an
``unfair distribution'', in that it is characterized by a majority
of poor agents and very few rich agents,
as signaled in particular by a zero mode
and by the exponential tail.
The form of distribution (\ref{BD}) will be referred to as the
Boltzmann distribution and is also known as Gibbs distribution.


\subsection{Minimum investment model without saving}
\label{minimuminvestment}

Despite the Boltzmann distribution is robust respect to the variation
of several parameters, the way it depends on the details of the trading
rule is subtle.
For instance, in the model studied in \rf{Chakraborti2002a},
the equilibrium distribution can have a very different shape.
In that model it is assumed that both economic agents $i$ and $j$
invest the same amount $ x_{min} $,
which is taken as the minimum wealth of the two agents,
$ x_{min} = \mr{min}\{x_i,x_j\}$.
The wealths after the trade are
$ x_{i}^{\prime }=x_{i} + \Delta x $ and
$ x_{j}^{\prime }=x_{j} - \Delta x $,
where $\Delta x = (2\epsilon -1) x_{min} $.

We note that once an agent has lost all his wealth, he is unable to trade
because $ x_{min} $ has become zero.
Thus, a trader is effectively driven out of the market
once he loses all his wealth. In this way,
after a sufficient number of transactions
only one trader survives in the market with the entire
amount of wealth, whereas the rest of the traders have zero wealth.
\begin{figure}[tb]
  \begin{center}
    \includegraphics[angle=0,width=.49\textwidth]{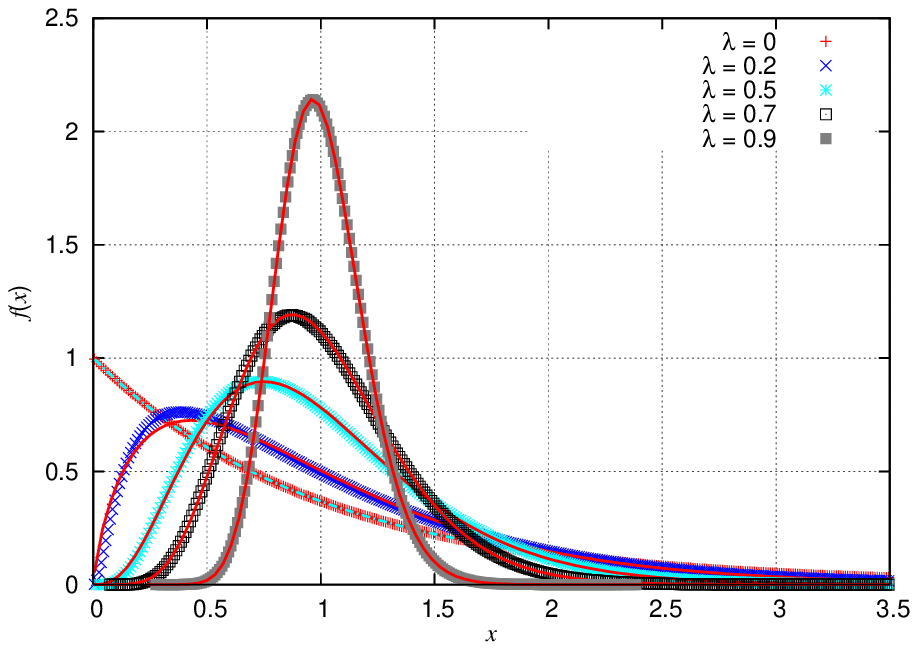}
    \includegraphics[angle=0,width=.49\textwidth]{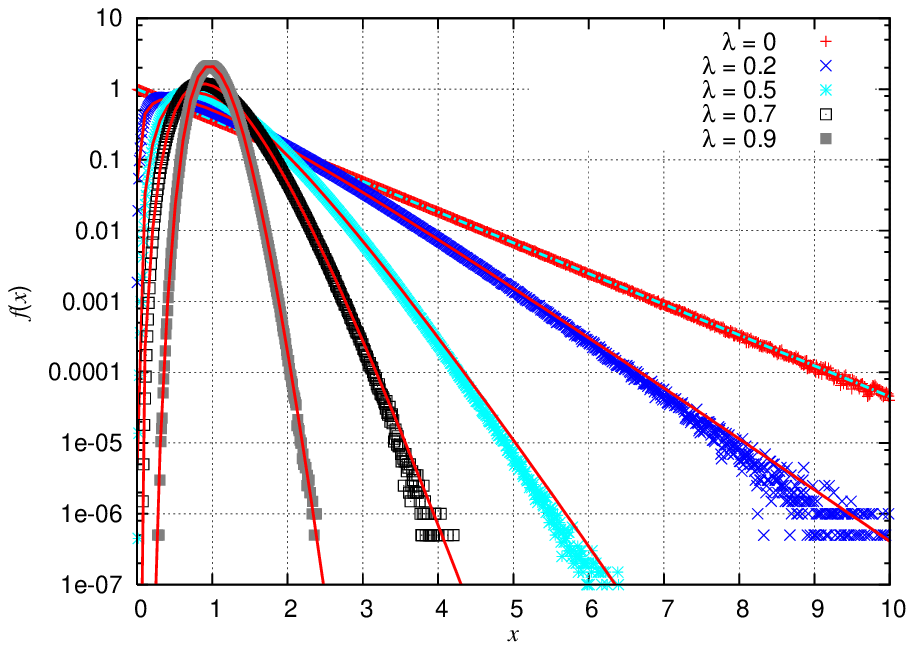}
    \caption{ Probability density for wealth $x$.
    The curve for $\lambda=0$ is the Boltzmann function
    $f(x)=\la x \ra^{-1}\exp(-x/\!\la x \ra)$ for the basic model
    of Sec.~\ref{sec:basic}.
    The other curves correspond to a global saving propensity
    $\lambda>0$, see Sec.~\ref{global}.}
    \label{fig:gamma}
  \end{center}
\end{figure}
%


\subsection{Model with constant global saving propensity: \\
{G}amma distribution}
\label{global}

A step toward generalizing the basic model and making it more realistic is
the introduction of a saving criterion regulating the trading dynamics.
This can be achieved defining a saving propensity
$0 < \lambda < 1$, that represents the fraction of wealth saved
--- and not reshuffled --- during a transaction.
The dynamics of the model is as follows
\cite{Chakraborti2000a,Chakraborti2002a}:
\begin{eqnarray}
  x_i' &=& \lambda x_i + \epsilon (1-\lambda) (x_i + x_j) \, ,
  \nonumber \\
  x_j' &=& \lambda x_j + \bar{\epsilon} (1-\lambda) (x_i + x_j) \, ,
  \label{sp1}
\end{eqnarray}
with $\bar{\epsilon} = 1 - \epsilon$,
corresponding to a $\Delta x$ in \eq{basic0} given by
\begin{equation}
  \Delta x
  =     (1 - \lambda) [ \bar{\epsilon} x_i - \epsilon x_j ]  \, .
\end{equation}
This model leads to a qualitatively different equilibrium distribution.
In particular, it has a mode $x_m>0$ and a zero limit for small $x$,
i.e.\ $f(x \! \to 0) \! \to 0$, see \fg{fig:gamma}.
The functional form of such a distribution
has been conjectured to be a Gamma distribution
on the base of an analogy with the kinetic theory of gases,
which is consistent with the excellent fitting provided
to numerical data \cite{Patriarca2004b,Patriarca2004c}.
Its form can be conveniently written
by defining the effective dimension $D_\lambda$ as \cite{Patriarca2004c}
\begin{equation}
  \frac{D_\lambda}{2}
  = 1 + \frac{3 \lambda}{1 - \lambda}
  = \frac{1 + 2\lambda}{1 - \lambda} \, .
  \label{D}
\end{equation}
According to the equipartition theorem, one can introduce
a corresponding temperature defined by the relation
$\la x \ra = {D_\lambda} T_\lambda / 2$, i.e.
\begin{equation}
  T_\lambda
  = \frac{2\la x \ra}{D_\lambda}
  = \frac{1 - \lambda}{1 + 2\lambda}\la x \ra
  \, .
  \label{T}
\end{equation}
Then the distribution for the reduced variable
%
$ \xi = x/T_\lambda$
%
reads
\begin{equation}
  f(\xi)
  = \frac{1}{\Gamma(D_\lambda/2)} \, \xi^{D_\lambda/2-1} \exp( - \xi )
  \equiv \gamma_{D_\lambda/2}(\xi) \, ,
  \label{Pxi}
\end{equation}
i.e.\ a Gamma distribution of order $D_\lambda/2$.
For integer or half-integer values of $n=D_\lambda/2$, this function
is identical to the equilibrium Maxwell-Boltzmann distribution
of the kinetic energy for a system of molecules in
thermal equilibrium at temperature $T_\lambda$
in a $D_\lambda$-dimensional space (see Appendix A for a detailed derivation).
For $D_\lambda = 2$, the Gamma distribution reduces to the Boltzmann distribution.
This extension of the equivalence between kinetic theory and closed economy models
to values $0 \le \lambda < 1$ is summarized in Table \ref{tab:analogy}.
\begin{table}
\centering
\caption{Analogy between kinetic and multi-agent model}
\label{tab:analogy}
\begin{tabular}{lll}
\hline\noalign{\smallskip}
                & Kinetic model       & Economic model  \\
\noalign{\smallskip}\hline\noalign{\smallskip}
variable        & $K$ (kinetic energy)& $x$ (wealth) \\
units           &  $N$ particles      & $N$ agents\\
interaction     & collisions          & trades\\
dimension       & integer $D$         & real number $D_\lambda$ [see \eq{D}]\\
temperature     & $\kB T=2\la K \ra/D$& $T_\lambda=2\la x \ra/D_\lambda$\\
reduced variable& $\xi = K/\kB T$     & $\xi = x / T_\lambda$\\
equilibrium distribution & $f(\xi)=\gamma_{D/2}(\xi)$ & $f(\xi)=\gamma_{D_\lambda/2}(\xi)$\\
\noalign{\smallskip}
\hline
\end{tabular}
\end{table}
This equivalence between a multi-agent system
with a saving propensity $0 \le \lambda < 1$ and an $N$-particle system
in a space with effective dimension $D_\lambda \ge 2$
was originally suggested by simple considerations
about the kinetics of a collision between two molecules.
In fact, for kinematical reasons during such an event only a fraction of
the total kinetic energy can be exchanged.
Such a fraction is of the order of $1-\lambda \approx 1/D$,
to be compared with the expression $1-\lambda = 3/(D/2+2)$
derived from \eq{D} \cite{Patriarca2004c}.
\begin{figure}[tb]
  \begin{center}
    \includegraphics[angle=0,width=.49\textwidth]{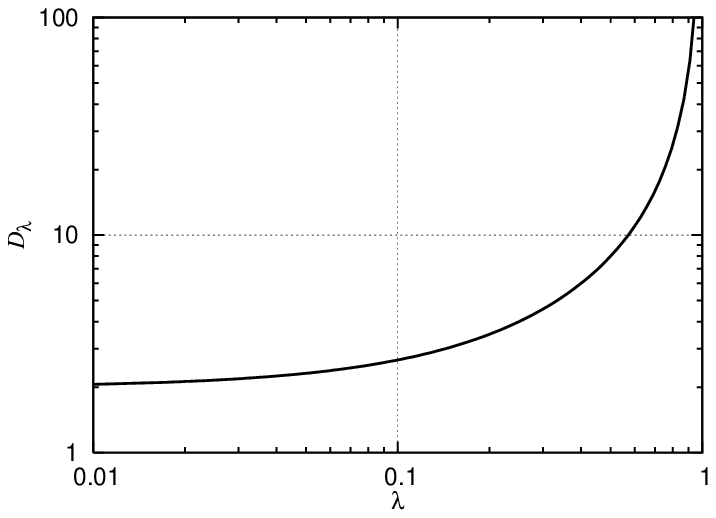}
    \includegraphics[angle=0,width=.49\textwidth]{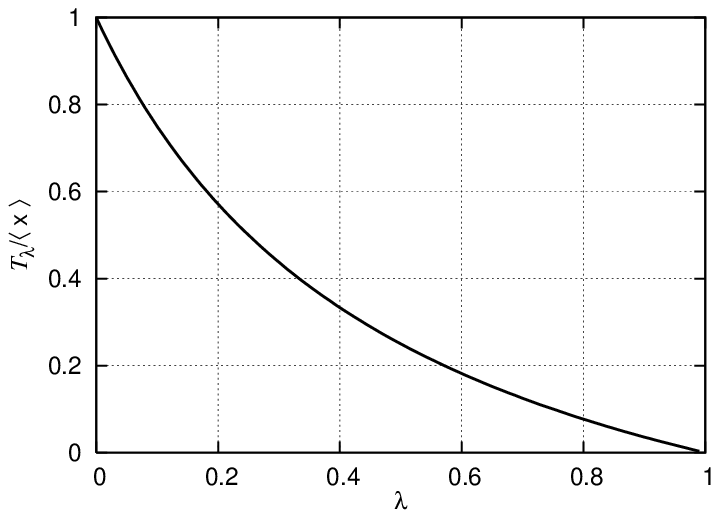}
    \caption{ Effective dimension $D_\lambda$, \eq{D},
    and temperature, \eq{T}, as a function
    of saving propensity $\lambda$.}
    \label{fig_Dlambda}
  \end{center}
\end{figure}
While $\lambda$ varies between 0 and 1, the parameter $D_\lambda$
monotonously increases from 2 to $\infty$,
and the effective temperature $T_\lambda$ correspondingly decreases
from $\la x \ra$ to zero; see \fg{fig_Dlambda}.
It is to be noticed that according to the equipartition theorem
only in $D_\lambda=2$ effective dimensions ($\lambda=0$)
the temperature coincides with the average value $\la x \ra$,
$T_\lambda = 2 \la x \ra/2 \equiv \la x \ra$,
as originally found in the basic model
\cite{Bennati1988a,Bennati1988b,Bennati1993a,Dragulescu2000a}.
In its general meaning, temperature represents rather an estimate
of the fluctuation of the quantity $x$ around its average value.
The equipartition theorem always gives a temperature smaller
than the average value $\la x \ra$ for a number of dimensions
larger than two.
In the present case, \eqs{D} or (\ref{T}) show that this happens
for any $\lambda>0$.

The dependence of the fluctuations of the quantity $x$
on the saving propensity $\lambda$ was studied in \rf{Chakraborti2000a}.
In particular, the decrease in the amplitude of the fluctuations
with increasing $\lambda$ is shown in  \fg{fig:dx}.

The fact that in general the market temperature $T_\lambda$
decreases with $\lambda$ means smaller fluctuations of $x$
during trades, consistently with the saving criterion,
i.e.\ with a $\lambda>0$.
One can notice that in fact
$T_\lambda = (1-\lambda)\la x \ra/(1+2\lambda) \approx (1-\lambda)\la x \ra$
is of the order of the average amount of wealth exchanged
during a single interaction between agents, see \eqs{sp1}.

\begin{figure}[tb]
  \begin{center}
  \resizebox{.75\textwidth}{!}
    {\includegraphics*[0cm,11cm][18cm,25cm]{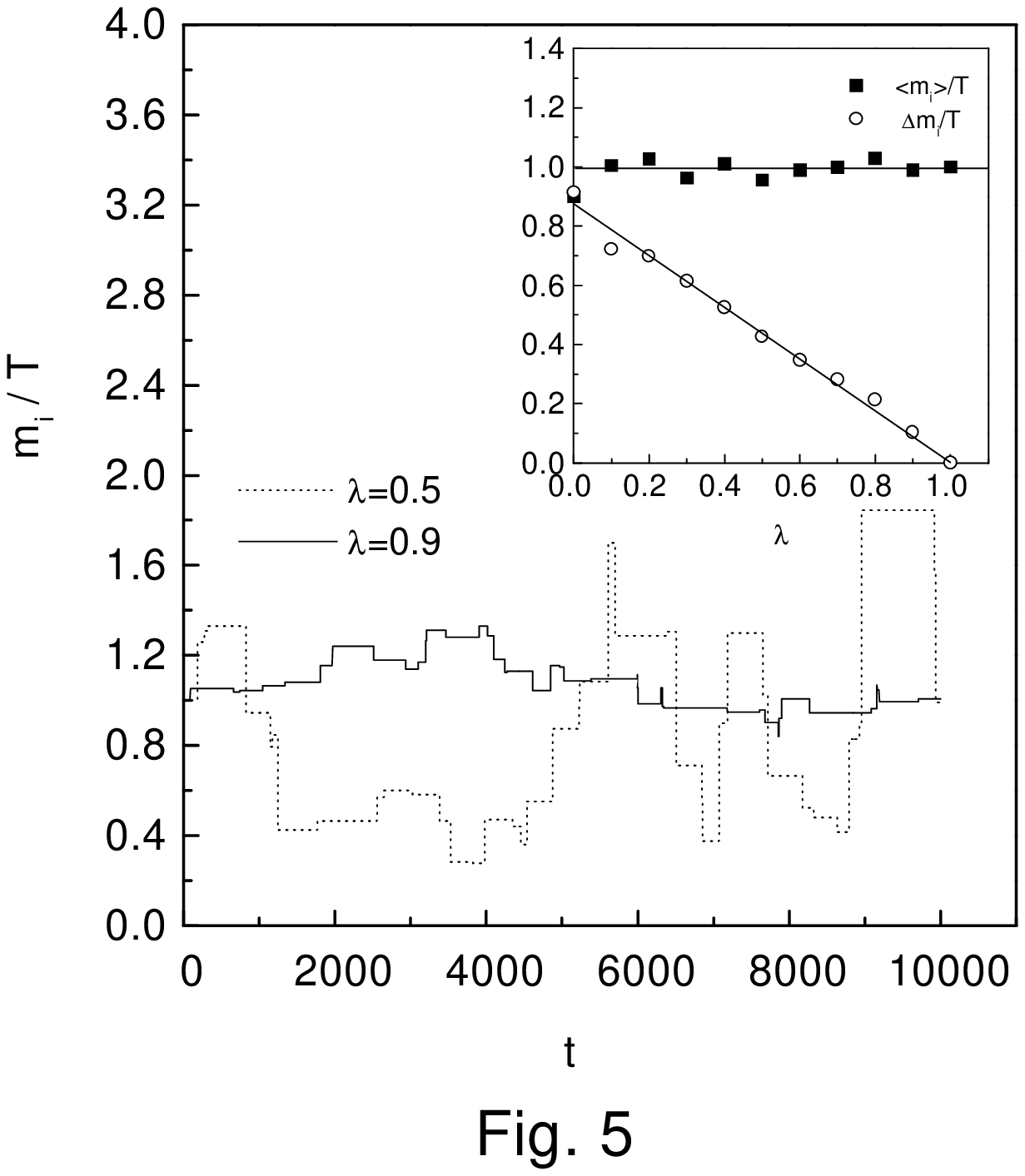}}
 \caption{ Reproduced from \rf{Chakraborti2000a} (only here $m \equiv x$
    while $T\equiv 1$ is a constant).
    The continuous and the dotted curves are the wealths of
    two agents with $\lambda=0.9$ and $\lambda=0.5$ respectively:
    notice the larger fluctuations in correspondence of the
    smaller $\lambda$.
    The inset shows that
    $\Delta m\equiv\Delta x=\sqrt{\la (x-\la x \ra)^2\ra}$
    decreases with $\lambda$.}
    \label{fig:dx}
  \end{center}
\end{figure}

\subsection{Model with individual saving propensities: Pareto tail}
\label{individual}

In order to take into account the natural diversity between various agents,
a model with individual propensities $\{\lambda_i\}$ as quenched random
variables was studied in \rfs{Chatterjee2003a,Chatterjee2004a}.
The dynamics of this model is the following:
\begin{eqnarray}
  x_i' &=&
  \lambda_i x_i + \epsilon [ (1-\lambda_i) x_i + (1-\lambda_j) x_j ] \, ,
  \nonumber \\
  x_j' &=&
  \lambda_j x_j + \bar{\epsilon} [(1-\lambda_i) x_i + (1-\lambda_j) x_j ] \, ,
  \label{sp2}
\end{eqnarray}
where, as above, $\bar{\epsilon} = 1 - \epsilon$.
This corresponds to a $\Delta x$ in \eq{basic0} given by
\begin{equation}
  \Delta x
  =  \bar{\epsilon} (1-\lambda_i) x_i - \epsilon (1-\lambda_j) x_j  \, .
\end{equation}

Besides the use of this trading rule,
a further prescription is given in the model,
namely an average over the initial random assignment of the
individual saving propensities:
With a given configuration $\{\lambda_i\}$,
the system is evolved until equilibrium is reached,
then a new set of random saving propensities $\{\lambda_i'\}$
is extracted and reassigned to all agents,
and the whole procedure is repeated many times.
As a result of the average over the equilibrium distributions
corresponding to the various $\{\lambda_i\}$ configurations,
one obtains a distribution with a power law tail,
$f(x) \propto x^{-\alpha-1}$, where the Pareto exponent
has the value $\alpha=1$.
This value of the exponent has been predicted by various theoretical
approaches to the modeling of multi-agent systems
\cite{Das2003a,Repetowicz2004a,Chatterjee2005a}.


\section{ Further analysis of the model with individual saving propensities}

On one hand, the model with individual saving propensities relaxes toward
a power law distribution --- with the prescription mentioned above
to average the distribution over many equilibrium states corresponding
to different configurations $\{\lambda_i\}$.
On the other hand, the models with a global saving propensity $\lambda>0$
and the basic model with $\lambda=0$, despite being
particular cases of the general model with individual saving propensities,
relax toward very different distributions,
namely a Gamma and a Boltzmann distribution,
respectively.
In this section we show that this difference can be reconciled
by illustrating how the observed power law is due
to the superposition of different distributions
with exponential tails corresponding to subsystems
of agents with the same value of $\lambda$.


\subsection{The $x$-$\lambda$ correlation}
\label{ correlation }

A key point which explains many of the features of the model
and of the corresponding equilibrium state
is a well-defined correlation
between average wealth and saving propensity,
which has been unnoticed so far in the literature \cite{Manna_here}.
The existence of such a correlation can be related to the origin of the power
law and its cut-off at high values of $x$. It also explains the paradox
according to which a very rich agent may lose all his wealth
when interacting with poor agents, as a consequence of
the stochastic character of the trading rule defined by \eq{sp2}.
\begin{figure}[tb]
  \begin{center}
    \includegraphics[angle=0,width=0.49\textwidth]{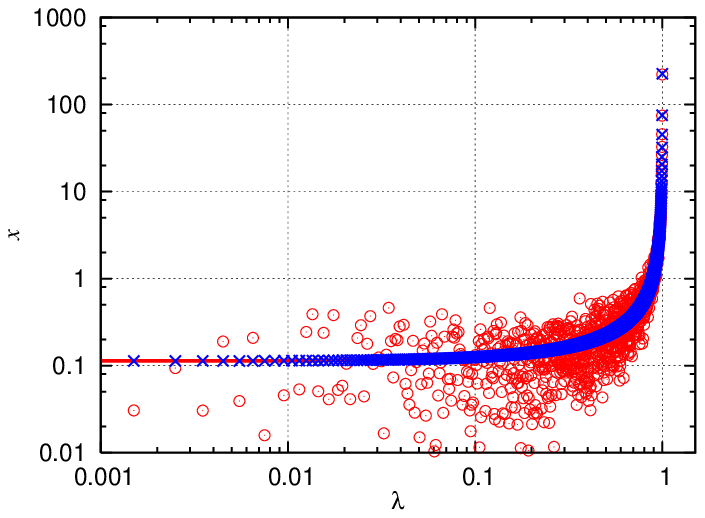}
    \includegraphics[angle=0,width=0.49\textwidth]{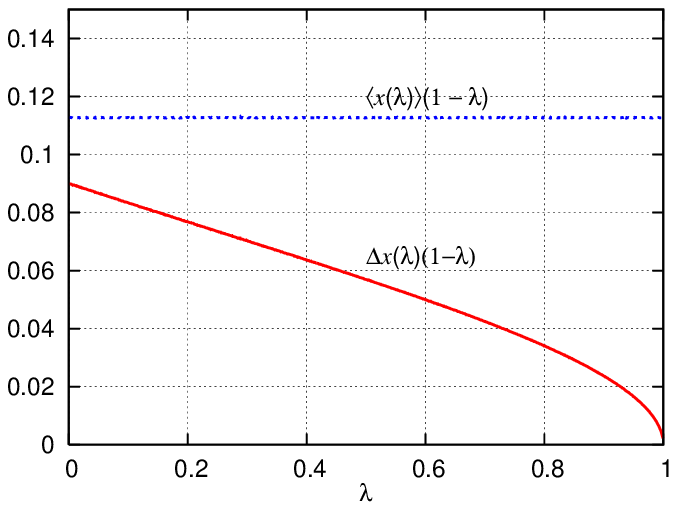}
    \caption{ Equilibrium state in the $x$-$\lambda$ plane
    after $t=10^9$ trades for a system of $N=1000$ agents.
    Left: Circles ($\circ$) represent agents,
    crosses ($\times$) represent the average wealth $\la x(\lambda) \ra$,
    the continuous line is the function
    $\la x(\lambda) \ra = \kappa/(1-\lambda)$, with $\kappa=0.1128$.
    Right: The product $\la x(\lambda) \ra (1-\lambda)$
    (dotted line) is constant, in agreement with \eq{x-lambda}.
    The product $\Delta x(\lambda) (1-\lambda)$ (continuous line),
    where $\Delta x(\lambda)$ is the standard deviation,
    shows that $\Delta x(\lambda)$ grows slower than $\la x(\lambda) \ra$.}
    \label{QL}
  \end{center}
\end{figure}
Figure \ref{QL} shows the equilibrium state 
for a system with $N=1000$ agents after $t=10^9$ trades.
Each agent is represented by a circle ($\circ$)
in the wealth-saving propensity $x_i$-$\lambda_i$ plane.
The correlation between wealth $x$ and saving propensity $\lambda$
becomes very high at large values of $x$ and $\lambda$.
Namely, one observes that
the average wealth $\la x(\lambda) \ra$
[crosses ($\times$) in \fg{QL}]
diverges for $\lambda \to 1$.
The average $\la x(\lambda) \ra$ was obtained by computing the probability
density $f(x,\lambda)$ in the $x$-$\lambda$ plane
(normalized so that $\int dx \, d\lambda \, f(x,\lambda)=1$)
and averaging for a fixed value of $\lambda$,
\begin{equation}
  \la x(\lambda) \ra = \int dx \, x f(x,\lambda) \, .
\end{equation}

The observed correlation naturally follows from the structure of the trade
dynamics (\ref{sp2}). We remind that initially every agent has the same
wealth $x_0=\la x \ra$. 
During the initial phase of the evolution, 
when all agents have approximately the same wealth $\la x \ra$, 
an agent $i$ with a large saving propensity
$\lambda_i$ can save more --- on average --- and therefore accumulate more.
Afterwards, the agent $i$ will continue to enter trades by investing only
a small fraction $(1-\lambda_i)x_i$ of his wealth $x_i$ in the trade.
Even when interacting with an agent $j$, with a smaller wealth $x_j<x_i$,
agent $i$ may still be successful in the trading, since agent $j$
may have also a smaller saving propensity $\lambda_j$,
so that the traded fraction of wealth $(1-\lambda_j)x_j$
is comparable with or even larger than $(1-\lambda_i)x_i$.
Trading by agent $i$ will very probably be successful on average 
with all agents $j$ with a $\lambda_j$ such that 
$(1-\lambda_j)x_j$ is smaller than $(1-\lambda_i)x_i$.
These considerations suggest that agent $i$
will reach equilibrium (and his maximum possible wealth) when
$(1-\lambda_i)x_i = \kappa \approx \la (1-\lambda)x\ra$.
The ratio between the constant $\kappa$ and
the average $\la (1-\lambda)x \ra = \sum_j (1-\lambda_j)x_j/N$
is actually found to be of the order of magnitude of 10.
The formula
\begin{equation}
  \la x(\lambda_i) \ra = \frac{\kappa}{1-\lambda_i} \, ,
  \label{x-lambda}
\end{equation}
however, shown as a continuous line in \fg{QL},
provides an excellent interpolation of the average wealth
$\la x(\lambda) \ra$ (also shown in the figure) computed numerically.


\subsection{ Variation of a single agent's wealth }

\begin{figure}[tb]
  \begin{center}
    \includegraphics[angle=0,width=0.72\textwidth]{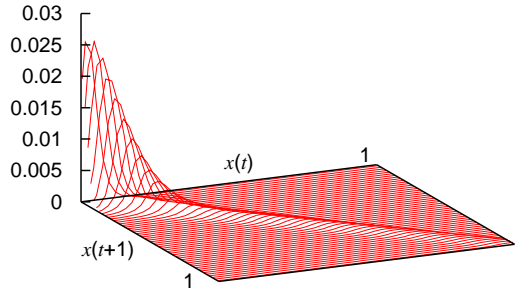}
    \vspace{-8mm}
    \caption{ Histogram of the wealths $x'=x(t+1)$ after a trade
    versus $x=x(t)$ before the trade for all agents and trades,
    in a system with $N=1000$ agents and $10^9$ trades.
    The distribution is narrower for large $x$ (rich agents),
    implying that it is unlikely that a rich agent becomes
    poor within a single trade.}
    \label{xx}
  \end{center}
\end{figure}
The stability of the asymptotic state is also shown by
the histogram in \fg{xx} of the wealths $x' \equiv x(t+1)$ 
after a trade versus $x \equiv x(t)$ before it defined in \eqs{basic0}.
The distribution is narrower at larger values of $x$ than at smaller ones,
implying that the probability that an agent $i$ will undergo
a large relative variation of his wealth $x_i$
within a single trade is much higher for poor agents.
The situation at small $x$ (corresponding to agents with smaller saving
propensities) is instead more similar to that of the trading rule
without saving ($\lambda=0$), \eqs{sp1}: the distribution is broader,
indicating a higher probability of a large wealth reshuffling during a trade.


\subsection{ Power laws at small $x$ and $t$ scales }
\label{ smallscale }

A peculiarity of the model with individual saving propensities
is noteworthy.
On one hand, in the procedure used to obtain a power law
in \rf{Chatterjee2003a} all agents are equivalent to each other:
they enter the dynamical evolution law on an equal footing,
their saving propensities are reassigned randomly
with the same uniform distribution between 0 and 1,
and even their initial conditions can be set to be
all equal to each other, $x_i = \la x \ra$, without loss of generality.
Therefore the various equilibrium configurations, corresponding to
different sets $\{\lambda_i\}$, are expected to be statistically equivalent
to each other, in the sense that one should be able to obtain the power law
distribution by a simple ensemble average for any fixed configuration of saving
propensities $\{\lambda_i\}$, if the number of agents $N$ is large enough.
On the other hand, an averaging procedure over several $\{\lambda_i\}$
configurations is in practice necessary to obtain a power law distribution.

In order to understand this apparent paradox, we checked how
the equilibrium distributions, corresponding to a given set of
saving propensities $\{\lambda_i\}$, look like.
One finds that every configuration $\{\lambda_i\}$ produces
equilibrium distributions very  different from each other;
see \fg{examples} for some examples.
The structures observed are very different from power laws,
with well resolved peaks at large $x$.
Only when an average over different $\{\lambda_i\}$ configurations
is carried out, one obtains a smooth power law
with Pareto exponent $\alpha=1$.
\begin{figure}[tb]
  \begin{center}
    \includegraphics[angle=0,width=0.45\textwidth]{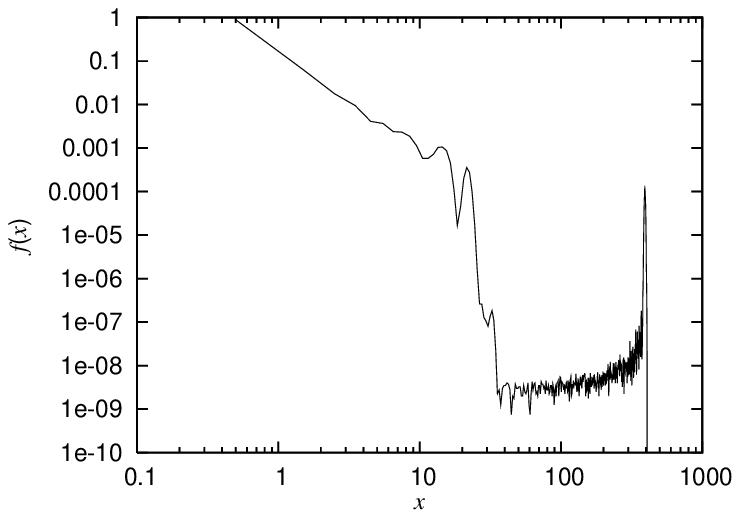}
    \includegraphics[angle=0,width=0.45\textwidth]{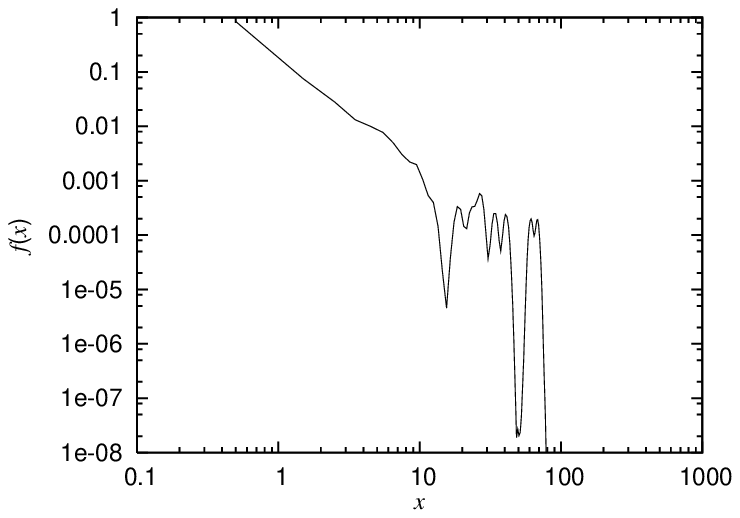}\\
    \includegraphics[angle=0,width=0.45\textwidth]{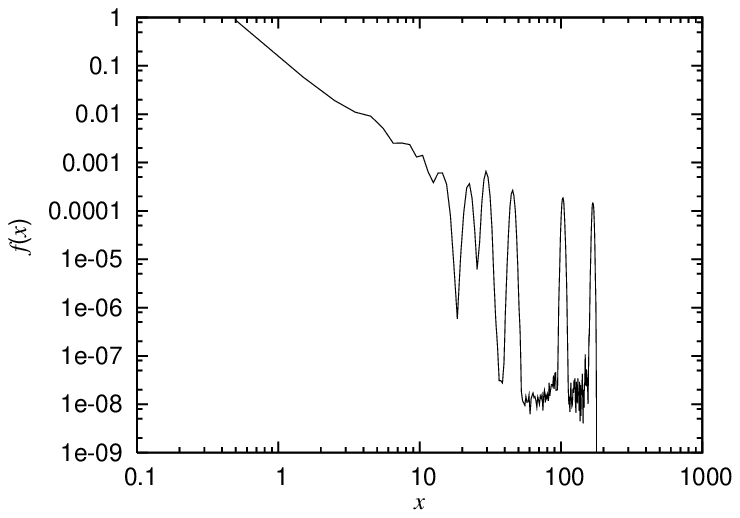}
    \includegraphics[angle=0,width=0.45\textwidth]{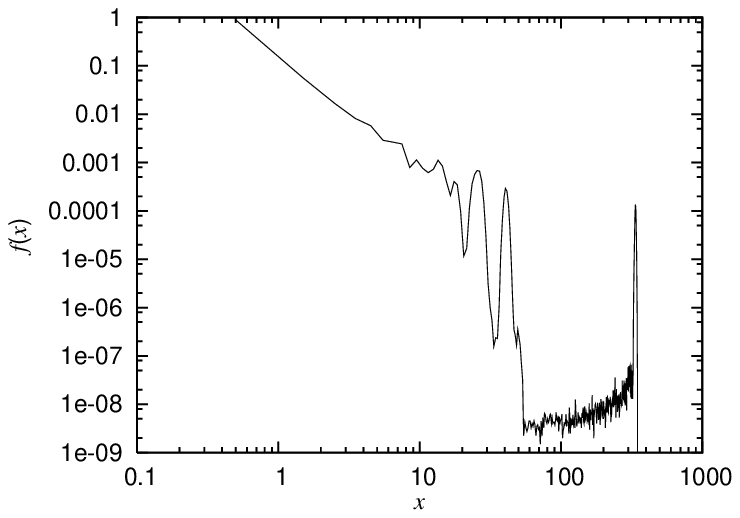}
    \caption{The equilibrium configurations corresponding to four
    different random saving propensity sets $\{\lambda_i\}$,
    for a system with $N=1000$ agents,
    differ especially at large $x$ where the distribution
    deviates from a power law.}
    \label{examples}
  \end{center}
\end{figure}
These same figures show, however, that for a given configuration
$\{\lambda_i\}$ a power law is actually observed at small values of $x$.
Another related interesting feature of simulations employing
a single saving propensity configuration $\{\lambda_i\}$
is that a power law distribution is found only on a limited time scale,
while it disappears partly or totally at equilibrium.
Thus also in the time dimension one surprisingly finds a distribution much
more similar to a power law at a smaller rather than larger scale.
\begin{figure}[tb]
  \begin{center}
    \includegraphics[angle=0,width=0.45\textwidth]{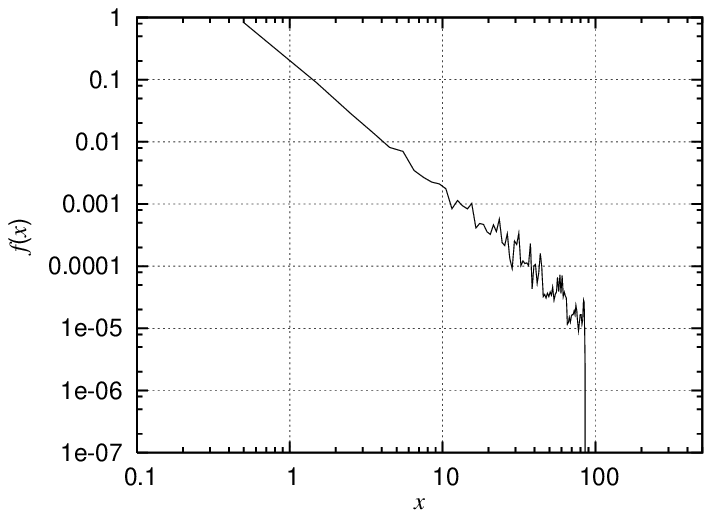}
    \includegraphics[angle=0,width=0.45\textwidth]{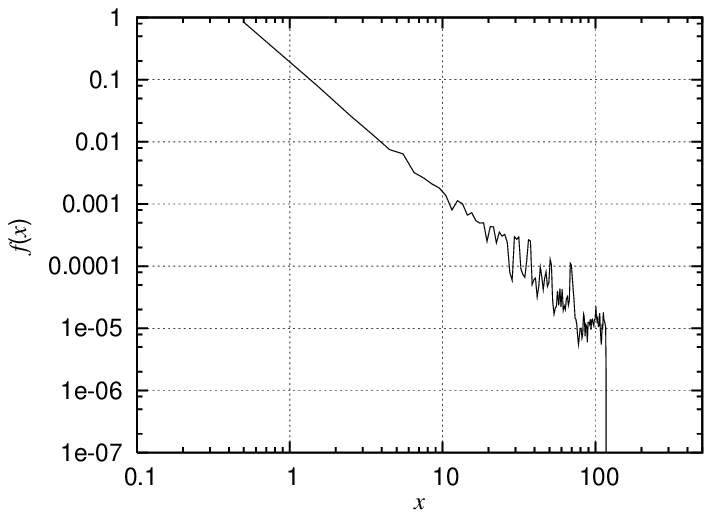}\\
    \includegraphics[angle=0,width=0.45\textwidth]{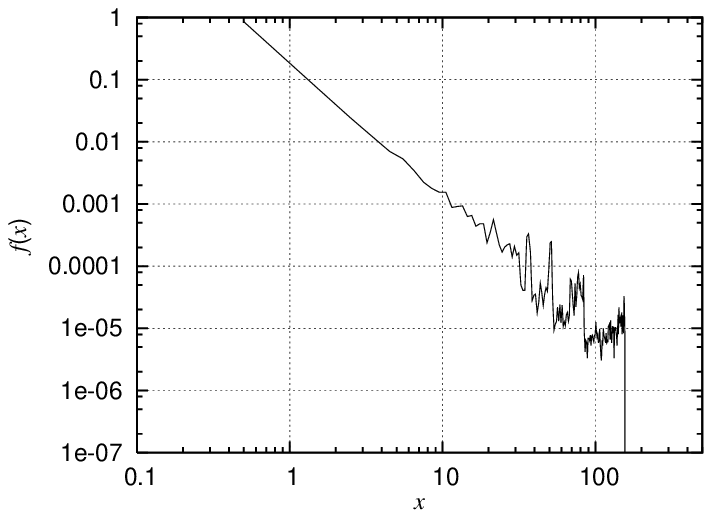}
    \includegraphics[angle=0,width=0.45\textwidth]{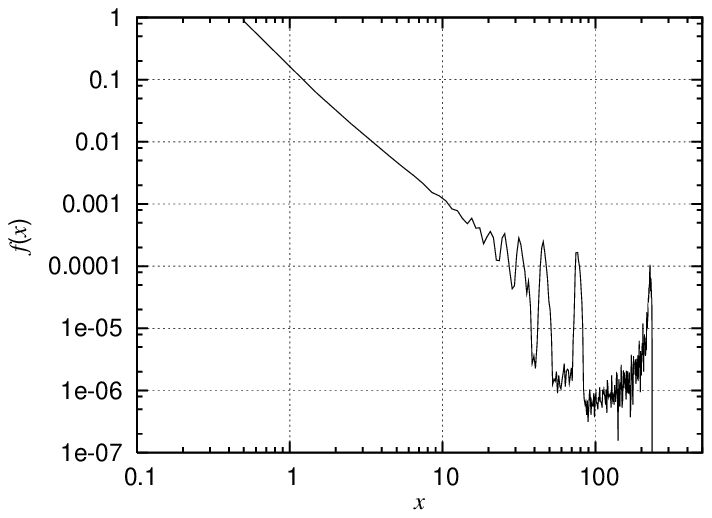}
    \caption{Time evolution of the $x$ distribution of a system
    with $N=1000$ agents:
    $t=2 \times 10^6$ (top left),
    $3 \times 10^6$ (top-right),
    $5 \times 10^6$ (bottom-left),
    and $2 \times 10^7$ trades (bottom-right).
    The distribution looks as a power law at small times,
    but develops into a structured distribution,
    maintaining a power law shape only at small $x$.}
    \label{evolution}
  \end{center}
\end{figure}
This is shown in the example in \fg{evolution}, where the distributions
of a system of 1000 agents at four different times are compared to each other.
These features suggest that the power law is intrinsically built into the
dynamical laws of the model but that, for some reasons,
it fades away at large $x$ and $t$ scales.
The $x$-$\lambda$ correlation discussed above in Sec.~\ref{ correlation }
can provide an explanation of these features, both for those in the $x$ and
in the time dimension, as discussed below.


\subsection{ Origin of the power law }
\label{ origin }

The peculiar features illustrated above,
the necessity of averaging over different configurations $\{\lambda_i\}$
as done in \rf{Chatterjee2003a} to obtain a power law distribution,
as well as the power law distribution itself,
are here explained in terms of equilibrium states of suitably
defined subsystems and the $x$-$\lambda$ correlation
illustrated above.
This may seem odd since at first sight the averaging procedure
of \rf{Chatterjee2003a} defines a nonequilibrium process,
the system being brought out of equilibrium
from time to time by the reassignment of the saving propensities.
Correspondingly, one may attribute the power law to the underlying
dynamical process, as it is often the case in nonequilibrium models
(e.g. models of markets on networks \cite{Chatterjee_here}).

\begin{figure}[tb]
  \begin{center}
    \includegraphics[angle=0,width=0.48\textwidth]{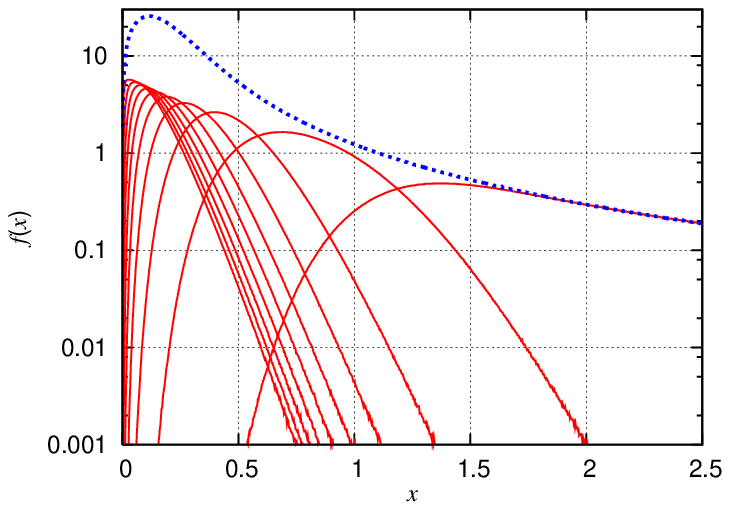}
    \includegraphics[angle=0,width=0.48\textwidth]{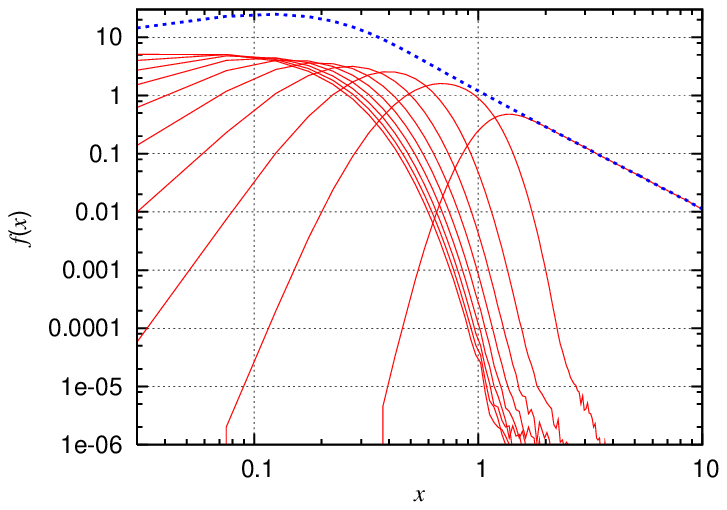}\\
    \includegraphics[angle=0,width=0.48\textwidth]{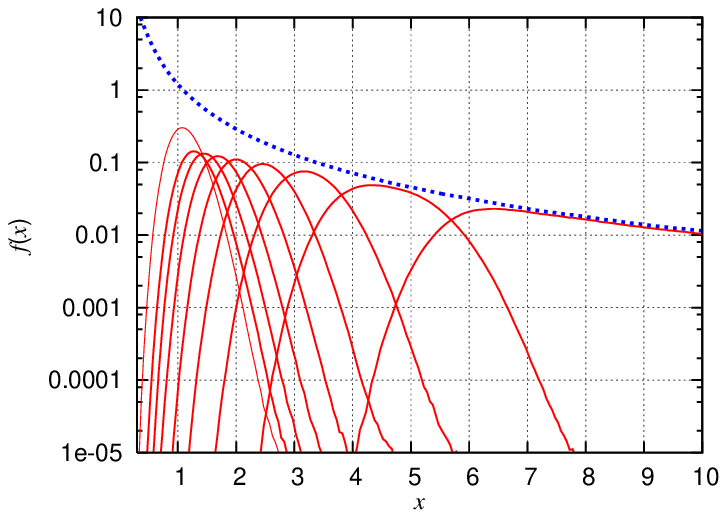}
    \includegraphics[angle=0,width=0.48\textwidth]{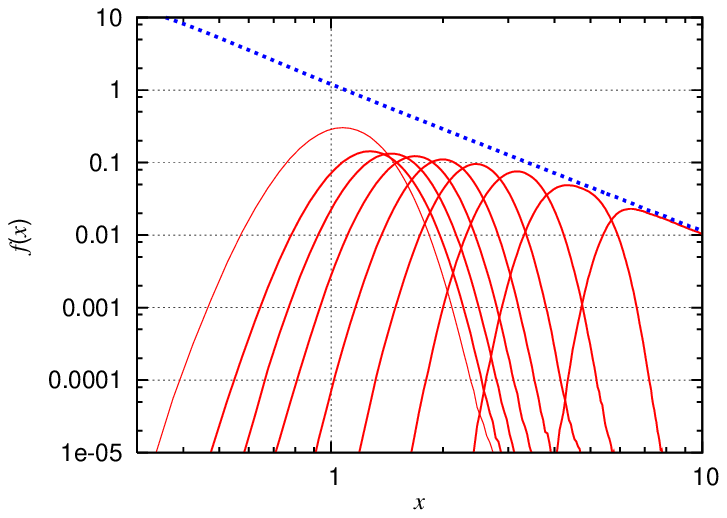}
    \caption{Semi-log and log plots of partial distributions
    (continuous curves) and the resulting overlap (dotted line).
    Above: Partial distributions from the 10 intervals of width
    $\Delta\lambda=0.1$ of the total $\lambda$ range (0,1).
    Below: The last partial distribution (with a power law tail) above,
    corresponding to the interval $\lambda=(0.9,1.0)$, has been further
    resolved into ten partial distributions for the sub-intervals of width
    $\Delta\lambda=0.01$.}
    \label{partials_a}
  \end{center}
\end{figure}

However, if one considers the partial distributions
of agents with a certain value of $\lambda$,
one finds an unexpected result.
For numerical reasons we consider the subsets made up of those
agents with saving propensity $\lambda$
within a window $\Delta\lambda$ around a given value $\lambda$.
Figure~\ref{partials_a} (upper row) shows the partial distributions
(continuous lines) of the ten subsystems
obtained by dividing the $\lambda$ range (0,1)
into ten slices of width $\Delta\lambda=0.1$
and average values 0.05,$\dots$,0.95 (curves from left to right respectively).
Most of the partial distributions have an exponential tail, and
only when summed up their overlap (dotted line)
reproduces a power law.
It can be noticed that the last partial distribution,
corresponding to the interval $\lambda=(0.9,1.0)$,
is not of exponential form, but rather presents a power law tail,
which overlaps with the total distribution at large $x$.
However, its power law form is due only to the low resolution
in $\lambda$ employed.
In fact even this partial distribution can in turn be shown
to be given by the superposition of exponential tails.
By increasing the resolution in $\lambda$,
i.e.\ using a smaller interval $\Delta\lambda=0.01$ to further resolve
the interval $\lambda=(0.9,1.0)$ into subintervals
with average values $\lambda=0.905,\dots,0.995$,
one obtains the partial distributions
shown in the lower row of \fg{partials_a}.
It is to be noticed that also in this case the last partial distribution
corresponding to the interval $\lambda=(0.99,1.00)$ has a power law tail.
The procedure can then in principle be reiterated to resolve also this
partial distribution by increasing the resolution in $\lambda$.

These facts also explain the origin of the peaks visible at large $x$ in the
plots in \fgs{examples} and \ref{evolution}.
Due to the high wealth-saving propensity correlation at large values of $x$,
see \fg{QL}, these peaks are due to agents with high $\lambda$.
The reason why these agents give rise to resolved peaks
instead of contributing to extending the power law tail is that
the partial distributions (i.e.\ the average values) of single agents
get farther and farther from each other for $\lambda \to 1$, while
the corresponding widths do not grow enough to ensure
the overlap of the distributions of neighbor agents in $\lambda$-space.
Eventually, each agent (or cluster of agents)
at high values of $x$ will be resolved as an isolated peak
against the background of the total distribution.
In greater detail, one finds that
the average value $\la x(\lambda) \ra$
diverges for $\lambda \to 1$ as $1/(1-\lambda)$,
as shown in \fg{QL}.
This implies that also the distance between
two generic consecutive agents increases:
if agents are labeled from $i=1$ to $i=N$
in order of increasing $\lambda$ ($\lambda_1 < \dots < \lambda_N$) and
the $\lambda$ distribution is uniform, then
$\Delta\lambda = \lambda_{i+1}-\lambda_i=\mr{constant}$.
The distance between the average positions of the partial distributions
of two consecutive agents is, from \eq{x-lambda},
\begin{equation}
  \delta\!\la x(\lambda) \ra
  = \la x(\lambda+\Delta\lambda) \ra -  \la x(\lambda) \ra
  \approx
  \frac{\pd \! \la x(\lambda) \ra}{\pd \lambda} \Delta\lambda
  \approx
  \frac{\kappa\Delta\lambda}{(1-\lambda)^2}   \, ,
\end{equation}
where $\kappa$ is a constant.
Thus $\delta\!\la x(\lambda) \ra$ diverges even faster
than $\la x(\lambda) \ra$.
At the same time, the width of the partial distribution $\Delta x(\lambda)$,
here estimated as
$\Delta x(\lambda)=\sqrt{\la x^2(\lambda)\ra-\la x(\lambda)\ra^2}$,
grows slower than $\la x(\lambda) \ra$, i.e.
for $\lambda \to 1$ the ratio
$\Delta x(\lambda)/\la x(\lambda) \ra \to 0$; see \fg{QL} (right).
The breaking of the power law and the appearance of the isolated peaks
takes place at a cutoff $x_c$ where the distance
$\delta\!\la x(\lambda) \ra$ between the peaks
corresponding to consecutive agents $i$ and $i+1$
becomes comparable with the peak width $\Delta x(\lambda)$.

Also the origin of the peculiarities in the time evolution
of the distribution function, mentioned in Sec.~\ref{ smallscale },
can now be explained easily.
In order to reach the asymptotic equilibrium state,
agents can rely on an income flux which is on average
proportional to $x_i(1-\lambda_i)$.
At the beginning, when all agents have the same wealth $x_i = x_0$,
this quantity is smaller for agents with a larger $\lambda_i$;
and with this smaller flux agents with large $\lambda_i$
have to reach their higher
asymptotic value $ \propto 1/(1-\lambda_i)$.
As a consequence, the relaxation time of an agent
is larger for larger $\lambda$,
a result already found in the numerical simulations
of the multi-agent model with fixed global saving propensity
(see Fig.~2 in \rf{Chakraborti2000a}).
Correspondingly, partial distributions of rich agents
will reach their asymptotic form later
(last frame in \fg{evolution}), while,
at intermediate times, their distributions will be spread
at smaller values of $x$, contributing to smoothen
the total distribution (first frame in \fg{evolution}).

It is also possible to explain why the averaging procedure
of \rf{Chatterjee2003a} is successful in producing a power law distribution.
Averaging over different configurations $\{\lambda_i\}$
is equivalent to simulate a very dense distribution in $\lambda$
--- which has large relaxation time and number of agents ---
with an affordable number of agents and computer time.
However, the procedure is not needed in principle,
since the power law can be obtained
also when a single configuration with a proper density
in $\lambda$-space is used.


\subsection{Checking different $\lambda$ distributions}
\label{fL}

\begin{figure}[tb]
  \begin{center}
    \includegraphics[angle=0,width=0.45\textwidth]{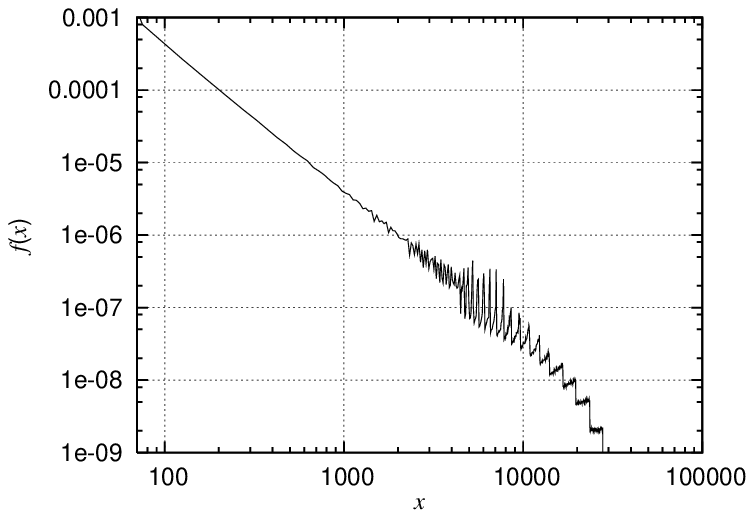}
    \includegraphics[angle=0,width=0.45\textwidth]{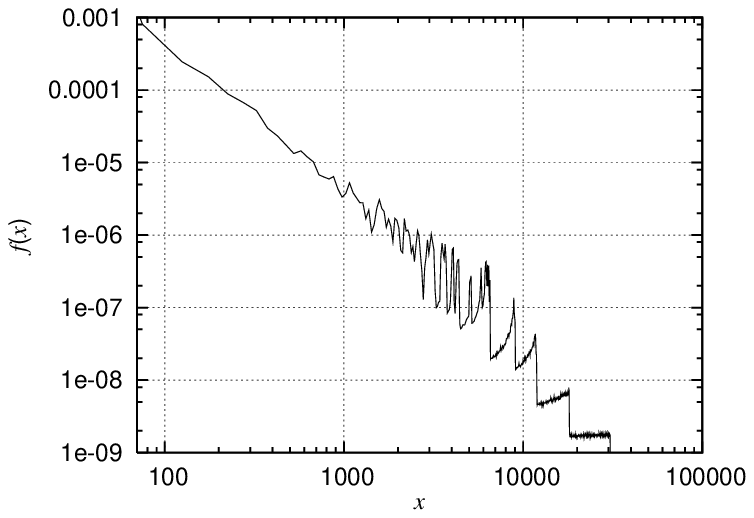}
    \caption{Wealth distribution of a system
    of $N=10^6$ agents after $10^{12}$ trades:
    the uniform $\lambda$ distribution \eq{Li} produces
    a smoother distribution with a power law shape extending to higher $x$
    (left) than for a random $\lambda$ distribution (right).}
    \label{Ldistr}
  \end{center}
\end{figure}
A practical way to avoid the appearance of the peaks at large $x$
and obtain a distribution closer to a power law is to increase the
density of agents, especially at values of $\lambda$ close to 1.
In a random extraction of $\{\lambda_i\}$, it is natural
that consecutive values of $\lambda_i$ will not be equally spaced.
Even small differences will be amplified
at large $x$ and will result in the appearance of peaks.
A deterministic assignment of the $\lambda$, e.g. a uniform distribution
achieved through the following assignment,
\begin{equation}
  \lambda_i = \frac{i}{N},~~~~i=0,N-1,
  \label{Li}
\end{equation}
is a uniform distribution of $\lambda$ in the interval [0,1)
and will generate a smoother distribution of $x$.
The comparison of the results for this distribution with those for a random
distribution of $\lambda$ is done in \fg{Ldistr} (notice the high value of
$N$). In the uniform case not only the power law extends
to higher values of $x$ but also that the distribution of peaks at large $x$
is globally smoother, in the sense that on average the single peaks follow
a power law better.


\section{Conclusions}
\label{conclusions}

We have reviewed some multi-agent models for the distribution
of wealth, in which wealth is exchanged at random in the presence of saving
quantified by the saving propensity $\lambda$.
We have shown how a distribution of $\lambda$ generates a power
law distribution of wealth through the superposition of Gamma distributions
corresponding to particular  subsets of agents.
The physical picture for the model with individual saving propensities
is thus more similar to that of the model with a constant global saving
propensity than it may seem at first sight.
In fact any subset of agents with the same value
of the saving propensity $\lambda$ equilibrates in a way similar
to agents in the model with a global saving propensity,
i.e.\ leading to a wealth distribution with an exponential tail.
Correspondingly we have shown that both the noise in the power law tail
and the cutoff in the power law depend on the coarseness of the $\lambda$
distribution. This extends the analogy between economic and gas-like systems
beyond the case of a global $\lambda \ge 0$,
characterized by a Maxwell-Boltzmann distribution,
to uniform continuous distributions in $\lambda$
that span the whole interval $\lambda \in [0,1)$.

\section*{Acknowledgment}

This work was partially supported by the Academy of Finland,
Research Centre for Computational Science and Engineering
project nr.~44897
(Finnish Centre for Excellence Program 2000-2005).
The work at Brookhaven
National Laboratory was carried out under Contract nr.\
DE-AC02-98CH10886, Division of Material Science,
U.S.\ Department of Energy.

\appendix

\section{Maxwell-Boltzmann distribution in $D$ dimensions}
\label{MB}

Here we show that for integer or half-integer values of
the parameter $n$ the Gamma distribution
\begin{equation}
  \gamma_n(\xi) = \Gamma(n)^{-1} \, \xi^{n-1} \exp(-\xi) \, ,
  \label{gamma}
\end{equation}
where $\Gamma(n)$ is the Gamma function,
represents the distribution
of the rescaled kinetic energy $\xi=K/T$
for a classical mechanical system in $D=2n$ dimensions.
In this section, $T$ represents the absolute temperature of the system
multiplied by the Boltzmann constant $\kB$.

We start from a system Hamiltonian of the form
\begin{equation}
  H(\mathbf{P},\mathbf{Q})
  = \frac{1}{2} \sum_{i=1}^N \frac{\mathbf{p}_i^2}{m_i} + V(\mathbf{Q}) \, ,
  \label{H}
\end{equation}
where $\mathbf{P}=\{\mathbf{p}_1,\dots,\mathbf{p}_N\}$ and
$\mathbf{Q}=\{\mathbf{q}_1,\dots,\mathbf{q}_N\}$
are the momentum and position vectors
of the $N$ particles,
while $V(\mathbf{Q})$ is the potential energy contribution
to the total energy.
For systems of this type, in which the total energy factorizes as a sum
of kinetic and potential contributions,
the normalized probability distribution in momentum space is simply
$f(\mathbf{P})=\prod_i (2\pi m_i T)^{-D/2} \exp(-{\mathbf{p}_i}^2/2m_i T)$.
Thus, since the kinetic energy distribution factorizes
as a sum of single particle contributions,
the probability density factorizes as a product of single particle
densities, each one of the form
\begin{equation}
  \label{p}
  f(\mathbf{p})
  =
  \frac{1}{(2\pi m T)^{D/2}}
  \exp\left(-\frac{{\mathbf{p}}^2}{2m T} \right) \, ,
\end{equation}
where $\mathbf{p}=(p_1,\dots,p_D)$ is the momentum of a generic particle.
It is convenient to introduce the momentum modulus $p$ of a particle in $D$ dimensions,
\begin{equation}
   p^2 \equiv \mathbf{p}^2  = \sum_{k=1}^{D} p_k^2  \, ,
\end{equation}
where the $p_k$'s are the Cartesian components,
since the distribution (\ref{p}) depends only
on $p \equiv \sqrt{ \mathbf{p}^2 }$.
One can then integrate the distribution over the $D-1$ angular variables
to obtain the momentum modulus distribution function,
with the help of the formula for the surface of a hypersphere
of radius $p$ in $D$ dimensions,
\begin{equation}
  S_D(p) = \frac{2\pi^{D/2}}{\Gamma(D/2)} \, p^{D-1} \, .
\end{equation}
One obtains
\begin{equation}
  f(p)
  = S_D (p) \, f(\mathbf{p})
  = \frac{2}{\Gamma(D/2)(2mT)^{D/2}} \, p^{D-1} \exp\left(-\frac{p^2}{2mT}\right) \, .
\end{equation}
The corresponding distribution for the kinetic energy $K=p^2/2m$ is therefore
\begin{equation}
  f(K)
  = \left[ \frac{dp}{dK} f(p) \right]_{p=\sqrt{2mK}}
  = \frac{1}{\Gamma(D/2) T}
    \left( \frac{K}{T} \right)^{D/2-1}
    \exp\left(-\frac{K}{T}\right) \, .
\end{equation}
Comparison with the Gamma distribution, \eq{gamma},
shows that the Maxwell-Boltzmann kinetic energy distribution in $D$ dimensions
can be expressed as
\begin{equation}
  f(K) = T^{-1} \gamma_{D/2}(K/T) \, .
\end{equation}
The distribution for the rescaled kinetic energy,
\begin{equation}
 \xi = K/T \, ,
\end{equation}
is just the Gamma distribution of order $D/2$,
\begin{equation}
  f(\xi)
  = \left[ \frac{dK}{d\xi}  \, f(K) \right]_{K=\xi T}
  = \frac{1}{\Gamma(D/2)} \, \xi^{D/2-1} \exp( - \xi )
  \equiv  \gamma_{D/2}(\xi) \, .
\end{equation}


 \bibliography{kolkata}
 \bibliographystyle{elsart-num}

\printindex
\end{document}